\documentclass[
  journal=proceedings,
  manuscript=article-type,
  year=2025
]{PMET_proc}

\usepackage{amsmath}
\newcommand{\bfs}[1]{\boldsymbol{#1}}   
\DeclareMathOperator{\Tr}{Tr}  
\usepackage[nopatch]{microtype}
\usepackage{booktabs}

\usepackage{orcidlink}
\newcommand{\orcidauthor}[2]{#1~\orcidlink{#2}}

\title{Detection of Cognitive Diagnostic Model Misspecification using New Lancaster-Chesher Information Matrix Tests}
\author{\orcidauthor{Richard M. Golden}{0000-0001-7505-6832}}   
\affiliation{COINS Lab, School of Behavioral and Brain Sciences, University
of Texas at Dallas, 75080, USA}
\email[Richard M. Golden]{golden@utdallas.edu}

\author{\orcidauthor{Reyhaneh Hosseinpourkhoshkbari}{0009-0000-9638-6814}}    
\affiliation{COINS Lab, School of Behavioral and Brain Sciences, University
of Texas at Dallas, 75080, USA}


\begin{document}

\begin{abstract}
Model specification tests play a crucial role in evaluating the appropriateness of
probability models for estimation and inference. Existing methods for the
detection of model misspecification such as the chi-square goodness-of-fit
(GOF) tests and more recently the M2 statistic 
\autocite{MaydeuOlivares2005,MaydeuJoe2014}
tend to
result in test statistics with excessive degrees of freedom for models with larger numbers 
of parameters. An alternative approach is based upon the Information Matrix (IM) equality. The IM equality
asserts that if a probability model is correctly
specified, the asymptotic covariance matrix of the maximum likelihood estimators can be
asymptotically estimated using a methodology based upon either the first or second derivatives
of the log-likelihood function. Using a contrapositive argument, White (1982)
\nocite{Wh82} proposed a
misspecification test methodology based upon comparing these two alternative covariance matrix
estimators. Extending this work, Presnell and Boos (2004) 
\nocite{Presnell2004}
showed how to develop a misspecification
test which only requires one degree of freedom regardless of the complexity of the model or data.
In this paper, we extend prior work and additionally apply methods of Golden et al. (2013, 2016) 
\nocite{golden2013}\nocite{Golden2016}
to derive and evaluate misspecification
tests for Cognitive Diagnostic Models (CDMs) which only require 1 or 2 degrees of freedom regardless
of model or data complexity. Analytic formulas for the tests are derived so they can be applied
without requiring computationally intensive
bootstrap simulation methods. Our simulation studies show the asymptotic statistical tests have good level (type 1
error) and power performance for CDM models and data which might be encountered in practice.
\end{abstract}

\section{Introduction}
A correctly specified probability model is capable of perfectly representing the data
generating process which generates observed data.
The detection of model misspecification is critical for evaluating the validity of 
model-based measurement processes. When model misspecification is present, the interpretation
of parameter estimates and the validity of inferences becomes questionable \autocite{hausman1978specification}.
Therefore, methods for the assessment of the presence of model misspecification are desirable.

A variety of methods have been developed for the purpose of the assessment of model
misspecification. For example, graphical residual diagnostic methods \autocite{FoxWeisberg2018} are an important tool but do not provide an explicit
procedure for supporting a hypothesis-driven decision process designed to determine whether
or not a model should be classified as misspecified. Chi-square goodness-of-fit (GOF) tests
(e.g., Hosmer et al., 1997) 
\nocite{Hosmer1997}
are commonly utilized
for probability models with categorical response variables. Such tests compare the predicted
probabilities of response categories to observed response frequencies in the observed data but
such tests are not applicable to continuous response variables. Moreover, the degrees
of freedom for the chi-square GOF test statistic increase as an exponential function of the number
of response categories. An extension of the chi-square GOF test statistic provides an improvement
in statistical power by only examining the first and second moments of the categorical response
distribution resulting in a chi-square GOF test statistic whose degrees of freedom
increase as a quadratic function of the number of response categories. In practice, the number
of response categories is relatively large  so both of these methodologies face the 
challenge of test statistics with large variance which would be expected to have poor statistical power.

Using a different approach, White (1982, 1994) 
and Golden et al. (2013, 2016)
\nocite{White1994}
\nocite{golden2013}
\nocite{Golden2016}
\nocite{White1994}
introduced a misspecification detection
methodology which, unlike chi-square GOF tests and graphical residual diagnostic methods,
is based upon testing the Information Matrix (IM) equality. The IM equality states that the
asymptotic covariance matrix of the maximum likelihood estimates can be estimated using either
the Hessian or Outer Product Gradient (OPG) covariance matrix estimators when the probability
model is correctly specified. Using a contrapositive argument, White (1982) noted that if
the Hessian and OPG covariance matrix estimators are asymptotically different, then this is
an indication of the presence of model misspecification. 

Assume a Data Generating Process (DGP) exists which consists of $n$
independent and identically distributed $d$-dimensional
random vectors (observations) 
$\tilde{\bf x}_1, \ldots, \tilde{\bf x}_n$ with common probability density
$p_e({\bf x})$. The observed data sample is a realization of 
$\tilde{\bf x}_1, \ldots, \tilde{\bf x}_n$ which will be denoted as
${\bf x}_1, \ldots, {\bf x}_n$.
Let $p({\bf x} | {\bfs \theta})$ denote a proposed
probability model specification for the DGP. If the probability model 
${\cal M} \equiv \{ p({\bf x} | {\bfs \theta}) : {\bfs \theta} \in {\cal R}^q \}$
is correctly specified, then there exists a $q$-dimensional parameter vector ${\bfs \theta}$ such
that $p({\bf x} | {\bfs \theta}) = p_e({\bf x})$ almost everywhere\footnote{The
"almost everywhere" condition is required when $\tilde{\bf x}_i$ is an absolutely
continuous or mixed random vector but can be omitted when $\tilde{\bf x}_i$ is a discrete
random vector.}

The maximum likelihood estimate (MLE) $\hat{\bfs \theta}_n$ is defined as a strict local
minimizer of the negative normalized log-likelihood function 
$\hat{\ell}_n({\bfs \theta}) \equiv -(1/n) \sum_{i=1}^n \log p({\bf x}_i | {\bfs \theta})$.
The Hessian covariance matrix estimator, $\hat{\bf A}_n$, is defined as
$\hat{\bf A}_n \equiv (1/n) \sum_{i=1}^n \hat{\bf A}^i$ where
$\hat{\bf A}^i$ is second derivative of $-\log p({\bf x}_i | {\bfs \theta})$ evaluated at 
$\hat{\bfs \theta}_n$.
The OPG covariance matrix estimator, $\hat{\bf B}_n$,
is defined as the inverse of the 
OPG matrix $(1/n) \sum_{i=1}^n \hat{\bf g}^i [\hat{\bf g}^i]^T$ where $\hat{\bf g}^i$ is the
transpose of the gradient of $-\log p({\bf x}_i | {\bfs \theta})$ evaluated at 
$\hat{\bfs \theta}_n$. 
Given appropriate regularity conditions, it can be shown that 
$\hat{\bf A}_n \rightarrow {\bf A}^*$ and $\hat{\bf B}_n \rightarrow {\bf B}^*$ with probability
one as $n \rightarrow \infty$.
If ${\bf A}^* \neq {\bf B}^*$, this implies the presence of model misspecification 
by the IM equality (White, 1982, 1987, 1994; Golden, 2020).

Presnell and Boos (2004) proposed that the comparison of $\hat{\bf A}_n$ and $\hat{\bf B}_n$
could be achieved by computing the "in-and-out-of-sample" IOS test statistic $s_{IOS}$.
Let ${\bfs \theta}_n(k)$ be the MLE obtained when the $k$th observation ${\bf x}_k$ is deleted
from the data sample. The negative normalized log-likelihood evaluated at $\hat{\bfs \theta}_n^-$
is denoted as $\hat{\ell}^-_n(\hat{\bfs \theta}_n^-) = (k-1)^{-1} \sum_{i\neq k} \log p({\bf x}_i | {\bfs \theta}_n(i))$ and corresponds to an "out of sample" estimate of model fit.
The IOS test statistic is then specified as:
\begin{equation}
\label{IOSteststatistic}
\hat{s}_{ios} =  -n\hat{\ell}_n(\hat{\theta}_n)  + n \hat{\ell}^-_n(\hat{\bfs \theta}_n^-).
\end{equation}
Presnell and Boos (2004) showed how this statistic is related to the IM testing methodology
of White (1982, 1994) by proving that $\hat{s}_{ios}$ converges in probability to the quantity
$\Tr([{\bf A}^*]^{-1} {\bf B}^*)$. If the model is correctly specified, then
${\bf A}^* = {\bf B}^*$ it then follows that $\Tr([{\bf A}^*]^{-1} {\bf B}^*) = q$.
In addition, Presnell and Boos (2004) showed that the asymptotic distribution
of the test statistic
$\hat{s}_{ios}$ was Gaussian corresponding to a one degree of freedom chi-square statistic.
Thus, Presnell and Boos (2004) provided the foundations of a
one degree of freedom misspecification detection methodology
for detecting model misspecification by testing the null hypothesis that
$\hat{s}_{ios} \rightarrow q$.

Golden et al. (2013, 2016) proposed a generalization of the White (1982) Information Matrix testing
framework termed the Generalized Information Matrix Test (GIMT) framework. The fundamental
concept of the GIMT framework involves comparing specific nonlinear functions of the Hessian
and OPG covariance matrices. Thus, the previous work by White (1982) and Presnell and Boos (2004)
can be viewed naturally as special cases of GIMTs. Golden et al. (2016) distinguishes between
two types of GIMTs. Let $H$ be a null hypothesis for a GIMT which means that if $H$ is false then
${\bf A}^* \neq {\bf B}$. If $H$ is formally equivalent to the null hypothesis that
${\bf A}^* = {\bf B}^*$ then $H$ is called {\em non-directional} otherwise $H$ is called 
{\em directional}.

In this paper we apply the GIMT theory
developed by Golden et al. (2013, 2016) to develop and evaluate one 
directional GIMT and one non-directional GIMT for the detection of model
misspecification.
The first specification test is a directional GIMT
which is closely related to the Presnell and Boos (2004) IOS test and the Golden et al. (2016) Robust Log
GAIC GIMT. In particular, a test statistic for testing the
null hypothesis that $\Tr([{\bf B}^*]^{-1} {\bf A}^*) = q$ is proposed.
This test which is called the {\em Inverse Trace GAIC} (InvTraceGAIC) test is shown using the methods
of Golden et al. (2016) to have a test statistic which is chi-square with only one degree of
freedom regardless of model or data complexity. The second specification test is a non-directional
GIMT whose null hypothesis is identical to the Composite Log GAIC GIMT described by Golden et al. (2016)
but we will refer to this test as the {\em trace inverse-trace GAIC} (TrInvTr) in this paper.

In order to explicitly derive a GIMT using the theory of Golden et al. (2013, 2016), a specific 
probability model must be specified because the GIMT Theory requires both the first and second
derivaties of the log-likelihood function. In this paper, the specified probability model
is a Cognitive Diagnostic Model (Rupp and Templin, 2008; George and Robitzsch, 2015).
Cognitive diagnostic models (CDMs) are a family of restricted latent class psychometric models designed to assess examinee skill mastery by incorporating prior knowledge of the 
relationship between latent skills and student responses.
\nocite{George2015}
\nocite{RuppTemplin2008}
In addition, a CDM 
outputs the attribute distribution or the probability that an examinee will have a particular set of 
skills given the examinee’s exam performance. Characteristics of latent skills in CDMs are defined 
through a Q-matrix that specifies which specific skills are relevant for answering a particular exam 
question. Furthermore, Q-matrix misspecification can affect parameter estimates and respondent 
classification accuracy (Rupp and Templin, 2008). In addition, the presence of model misspecification
can affect semantic interpretation of the unobservable latent skills which is crucial since CDMs are used
for diagnostic purposes. 

In the first part of the paper, we provide our theoretical contribution which shows how to
apply the theory of Golden et al. (2016) to develop these two new low degree
of freedom model misspecification tests for CDMs. The theoretical contribution includes new explicit
formulas which not only explicitly specify the new CDM-specific test statistics but also 
include estimates of their asymptotic covariance to support hypothesis testing. 

In the second part of the paper, we evaluate our new formulas in simulation studies involving
a CDM in which misspecified and correctly specified CDMs are fit to a known DGP. The discrimination
performance of the new GIMTs is reported using response operating curves (ROCs) and the performance
of the analytic formulas in estimating Type 1 error probabilities ("p-values") is evaluted using
p-value plots (*Davidson and MacKinnon, 1988).
\nocite{Davidson1998}

\section{Theoretical Results}
\label{theorysection}

\subsection{The New Generalized Information Matrix Tests}

The null hypothesis of a Generalized Information Matrix Test (GIMT) is specified by
\begin{equation}
\label{H0}
H_o : {\bf s}({\bf A}^*, {\bf B}^*) = {\bf 0}_r
\end{equation}
where the column vector-valued function ${\bf s}$ maps a $q$-dimensional matrix ${\bf A}$
and $q$-dimensional matrix ${\bf B}$ into an $r$-dimensional vector. The notation
${\bf 0}_r$ denotes an $r$-dimensional vector of zeros. 

Define the {\em selection test statistic} $\hat{\bf s}_n$ such that
$\hat{\bf s}_n \equiv {\bf s}(\hat{\bf A}_n, \hat{\bf B}_n)$.
Let $\hat{\bf C}^n_s$ specify an estimator of the asymptotic covariance of $\hat{\bf s}_n$.
To test the null hypothesis, the Wald test statistic 
\begin{equation}
\label{WaldTest}
\hat{\cal W}_n = n \hat{\bf s}_n^T [\hat{\bf C}^n_s]^{-1} \hat{\bf s}_n
\end{equation}
can be shown (Golden et al., 2016, Theorem 7) to have an asymptotic chi-square distribution
with $r$ degrees of freedom when $H_o$ in \ref{H0} is true. When $\ref{H0}$ is false,
the power of the statistical test converges to one when the sample size $n$ is sufficiently
large. The assumptions of this theorem require that
$\hat{\bf A}_n$ and $\hat{\bf B}_n$ are asymptotically positive definite.

Let the duplication matrix ${\cal D}_k : {\cal R}^{k(k+1)2} \rightarrow {\cal R}^{k^2}$ be defined such
that ${\cal D}_k {\bf vech}({\bf A}) = {\bf vec}({\bf A})$ and the 
inverse duplication matrix ${\cal D}_k^{\dagger}$ be defined 
such that ${\cal D}_k^{\dagger} {\bf vec}({\bf A}) = {\bf vech}({\bf A})$.
Let ${\cal D}_k^{\otimes} \equiv {\bf I}_2 \otimes {\cal D}_k$ where ${\bf I}_2$ is the 
two-dimensional identity matrix. 
Let ${\cal D}_k^{{\otimes}\dagger} = {\bf I}_2 \otimes {\cal D}_k^{\dagger}$.

Using the results of Golden et al. (2016),
the asymptotic covariance matrix of $\hat{\bf s}_n$, $\hat{\bf C}^n_s$ is given by the formula:
\begin{equation}
\label{GoldenEtAl16-EQ2}
\hat{\bf C}^n_s = (1/n) \sum_{i=1}^n \hat{\bfs \delta}_i \hat{\bfs \delta}_i^T
\end{equation}
where 
\begin{equation}
\label{part2GoldenEtAl16-EQ2}
\hat{\bfs \delta}_i = \nabla \hat{\bf s}_n {\cal D}_k^{\otimes}\left(
\hat{\bf d}_{i,n} - \ddot{\nabla} \hat{\bf d}_n [\hat{\bf A}_n]^{-1} \hat{\bf g}^i 
- \hat{\bf d}_n \right), \;
\nabla \hat{\bf s}_n \equiv 
\left[ \frac{d{\bf s}}{d{\bf A}}(\hat{\bf A}_n), \frac{d{\bf s}}{d{\bf B}}(\hat{\bf B}_n) \right],
\end{equation}
\begin{equation}
\label{GoldenEtAl16-EQ3}
\ddot{\nabla} \hat{\bf d}_n = {\cal D}_k^{{\otimes}\dagger}
\left[
\begin{array}{c}
\frac{d\hat{\bf B}_n}{d{\bfs \theta}} + n^{-1} \sum_{i=1}^n
vec(\hat{\bf A}_n - \hat{\bf B}_n) \hat{\bf g}_i^T \\
\frac{d\hat{\bf B}_n}{d{\bfs \theta}}
\end{array}
\right],
\end{equation}
and
\begin{equation}
\label{GoldenEtAl16-EQ4}
\frac{d\hat{\bf B}_n}{d{\bfs \theta}} = 
n^{-1} \sum_{i=1}^n \left[ \left( \hat{\bf A}_i \otimes \hat{\bf g}^i \right)
+ \left(\hat{\bf g}^i \otimes \hat{\bf A}_i \right) \right].
\end{equation}
Equations \ref{GoldenEtAl16-EQ2}, \ref{GoldenEtAl16-EQ3}, and
\ref{GoldenEtAl16-EQ4} correspond to Equations 2, 3, and 4
in Golden et al. (2016) respectively. Equations \ref{GoldenEtAl16-EQ3} and 
\ref{GoldenEtAl16-EQ4} specify the Lancaster-Chesher formula 
(Lancaster, 1984)
\nocite{lancaster1984covariance}
described in Golden
et al. (2016).

\subsubsection{Inverse-Trace GAIC}
The Inverse-Trace GAIC GIMT
is a statistical test for model misspecification which 
tests the null hypothesis 
\begin{equation}
\label{inversetracegaicHO}
H_o : {\bf s}({\bf A}^*, {\bf B}^*) = 0,  \;\;
{\bf s}({\bf A}^*, {\bf B}^*) = \log \left(q^{-1} [\Tr \left(({\bf B}^*)^{-1}{\bf A}^*\right) ] \right).
\end{equation}
When the  null hypothesis in (\ref{inversetracegaicHO}) is rejected, this
implies that ${\bf A}^* \neq {\bf B}^*$
which indicates model misspecification
by the IM Equality (White, 1982; Golden, 2020).
This null hypothesis and its associated GIMT are referred to as 
directional (see Golden et al., 2016) since rejection of $H_o$ implies
${\bf A}^* \neq {\bf B}^*$ but the converse of this statement does not
necessarily hold. The Inverse-Trace GAIC null hypothesis is similar but
not equivalent to the null hypotheses of the 
 IOS statistical test (Presnell and Boos) and the Robust Log GAIC
GIMT (Golden et al., 2016). 

The
selection statistic $\hat{\bf s}_n$ for the directional inverse-Trace GAIC 
test in (\ref{WaldTest}) is defined by the 
formula:
\begin{equation}
\label{SInvTraceGAIC}
\hat{\bf s}_n = \log \left(q^{-1} \Tr\left(\hat{\bf B}_n^{-1} \hat{\bf A}_n\right) \right).
\end{equation}
Since $\hat{\bf s}_n$ is a scalar, the test statistic $\hat{\cal W}_n$ in (\ref{WaldTest}) is a chi-square
random variable with one degree of freedom under the null hypothesis.

The derivative of ${\bf s}({\bf A}, {\bf B})$ in \ref{H_o} for the
inverse-trace GAIC is given explicitly by 
finding the analytic derivatives of ${\bf s}$ with respect to ${\bf A}$ and
${\bf B}$ which are
It can be show that evaluating the formula:
\begin{equation}
\label{DSInvTraceGAIC}
\nabla {\bf s}({\bf A}, {\bf B})
= \left[ \frac{d{\bf s}}{d{\bf A}}, \frac{d{\bf s}}{d{\bf B}} \right]
\end{equation}
at $\hat{\bf A}_n$ and $\hat{\bf B}_n$
where
\begin{displaymath}
\frac{d{\bf s}}{d{\bf A}} = 
\left(\Tr\left({\bf B}^{-1} {\bf A}\right)\right)^{-1}[{\bf vec}({\bf B}^{-1})]^T \;\;
\text{and} \;\;
\frac{d{\bf s}}{d{\bf B}} = 
-\left(\Tr\left({\bf B}^{-1} {\bf A}\right) \right)^{-1}
\left[{\bf vec}\left({\bf B}^{-1}{\bf A} {\bf B}^{-1}\right) \right]^T.
\end{displaymath}

\subsubsection{Trace Inverse-Trace GAIC}
The Trace Inverse-Trace GAIC GIMT
is a statistical test for model misspecification which 
tests the null hypothesis 
\begin{equation}
\label{traceinversetracegaicHO}
H_o : {\bf s}({\bf A}^*, {\bf B}^*) = {\bf 0}_2 \;\;\text{where}
\end{equation}
\begin{displaymath}
{\bf s}({\bf A}^*, {\bf B}^*) = 
\left[ 
\log \left(q^{-1} \Tr\left([{\bf A}^*]^{-1} {\bf B}^* \right) \right),
\log \left(q^{-1} \Tr\left([{\bf B}^*]^{-1} {\bf A}^* \right) \right)
\right]^T.
\end{displaymath}
When the  null hypothesis in (\ref{traceinversetracegaicHO}) is rejected, this
implies that ${\bf A}^* \neq {\bf B}^*$ which indicates model misspecification
by the IM Equality (White, 1982; Golden, 2020).
This null hypothesis and its associated GIMT are referred to as 
non-directional (see Golden et al., 2016) since 
it can be shown 
$H_o$ holds if and only if ${\bf A}^* = {\bf B}^*$ 
(Cho and Phillips, 2018; also see Golden, 2020, Chapter 16, Theorem 16.3.2).
The Trace Inverse-Trace GAIC null hypothesis is 
equivalent to the null hypothesis of the Composite Log GAIC
GIMT (Golden et al., 2016) although the associated test statistic is slightly
different. \nocite{ChoPhillips2018}

In Equation \ref{WaldTest}, the 
selection statistic $\hat{\bf s}_n$ for the non-directional trace inverse-Trace GAIC 
test is defined by the 
formula:
\begin{equation}
\label{SInvTraceGAIC}
\hat{\bf s}_n = 
\left[ 
\log \left(q^{-1} \Tr\left(\hat{\bf A}_n^{-1} \hat{\bf B}_n\right) \right),
\log \left(q^{-1} \Tr\left(\hat{\bf B}_n^{-1} \hat{\bf A}_n\right) \right)
\right]^T
\end{equation}
Since $\hat{\bf s}_n$ is a two-dimensional vector, 
the test statistic $\hat{\bf s}_n$ is a chi-square
random variable with two degrees of freedom under the null hypothesis.

\subsection{Cognitive Diagnostic Probability Model}
The Cognitive Diagnostic Probability model used in the mathematical derivations
and simulation studies corresponds to the model specified by
Hosseinpourkhoskbari and Golden (2024). For this model, the $j$th element of
${\bf x}_j$, $x_{ij}$, is the response of a student to an exam question which can take on the
value of either zero or one.
\nocite{HosseinpourGolden2024}

In particular,
\begin{equation}
\label{CDMprobabilitymodel}
p({\bf x}_i | {\bfs \theta}) = 
\sum_{ {\bfs \alpha}} p({\bf x}_i | {\bfs \alpha}, {\bfs \theta}) p({\bfs \alpha})
\end{equation}
where ${\bfs \alpha}$ is a $K$-dimensional vector consisting of binary latent
variables and the {\em proficiency model}
$p({\bfs \alpha})$ is a known probability of a particular ${\bfs \alpha}$.
Let $\theta_{j,1}$ and $\theta_{j,2}$ be free parameters
associated with the $j$th element of
${\bf x}_i$ respectively. Define 
${\bfs \theta} = [\theta_{1,1}, \theta_{1,2}, \ldots, \theta_{q/2,1}, \theta_{q/2,2}]$.
The {\em evidence model} is defined as:
\begin{displaymath}
p(x_{ij} = 1 | {\bfs \alpha}, {\bfs \theta}) = 
\left( 1 + \exp[-\theta_{j,1} \psi_j({\bfs \alpha} + \theta_{j,2}] \right)^{-1}.
\end{displaymath}

\section{Simulation Study}

\subsection{Data Set and Q matrix}
\nocite{Ta84}
The dataset extracted from
Tatsuoka (1984) consists of a set of math problems
designed to assess students’ ability to solve problems
involving fractions. In this case, we utilized the ’Fraction.1’ dataset from the ’CDM’
R package (George et al., 2016) containing the dichotomous responses of 536 middle
school students over 15 fraction subtraction test items. The Q-matrix specifying
which of five distinct skills were required to answer a particular test item was based
upon the Q-matrix used by Tatsuoka (1984) and provided in George et al. (2016).

\subsection{Methods}

\subsubsection{Methods: Type 1 Error Analytic Formula Performance}
\label{pvaluemethod}
The CDM was fit to the original data set with a sample size of 
$n = 536$ by 
minimizing the negative normalized log-likelihood function. Then 
500 simulated data sets each consisting of responses from $n=134$ participants
were generated from the fitted CDM. Next, the Lancaster-Chesher Wald test formulas whose derivations
were presented in Section \ref{theorysection} were used to detect misspecification for the Inverse-Trace GAIC
and Trace Inverse-Trace GAIC GIMTs for a wide range of significance levels. A p-value plot
was then generated which plots the percentage of times the GIMT incorrectly rejected the 
null hypothesis of correct specification as a function of the significance level of the test.

\subsubsection{Methods: Power Analyses using Alternative Models with Missing Skills}
\label{rocmethod}
Three different ("misspecified")
versions of the original fitted CDM used in the 
Type 1 Error Analyses (Section \ref{pvaluemethod})
were then defined by modifying
the Q-matrix in Section \ref{pvaluemethod}
in three different ways.
Specifically, the Q matrix in the original CDM
used in the p-value analyses was modified by: 
(1) deleting skill 5, 
(2) deleting skills 2 and 4, and (3) deleting skills 1,3, and 4. These three different Q matrices corresponded to 3 alternative CDMs which will be
referred to as ''misspecified models'' since they will be fit to data
generated by the original CDM used in Section \ref{pvaluemethod}.

The specific Q-matrix modifications were intentionally selected to represent different magnitudes of practically meaningful model misspecification. These changes were chosen intentionally to represent mild, moderate, and severe levels of misspecification.

In the original Q-matrix, Attribute 5 appears in only two items and plays a relatively small role in the assessment. Therefore, deleting Attribute 5 represents a mild form of misspecification. In contrast, Attributes 2 and 4 are mainly associated with the most complex items and are important for distinguishing higher-level mastery patterns. Removing both of these attributes leads to a much stronger distortion of the skill structure.

The third case—deleting Attributes 1, 3, and 4—represents an intermediate level of misspecification. Although Attributes 1 and 3 contribute to many items, the remaining attributes still preserve part of the original structure.

Together, these three conditions create a spectrum of misspecification severity, from weak to strong. This design allows us to evaluate the performance of the GAIC-based tests across a realistic range of situations where an attribute is unintentionally
omitted from the model.

The three misspecified CDMs and the 
original CDM used in the p-value analyses were then 
used to evaluate the discrimination performance of the
Inverse-Trace GAIC and Trace Inverse-Trace GAIC Wald tests
for misspecification detection. These results are reported
using ROC curves. In particular, 500 pairs of 
simulated data samples of $n=134$ data records were generated from the original CDM which had been fit to $n=536$ data records.

The first member of each pair was fit to the original CDM ("correctly specified model")
and the second member of each pair was fit to either the original CDM or a misspecified
model. The two GIMTs were then used to classify the fitted model as correctly specified
or misspecified for a wide range of significance levels. 

These results were then reported using
ROC curve plots where 
the true positive rate is plotted against the false positive rate for a range of different decision
thresholds (i.e., significance levels).
Additionally, the Area Under the ROC (AUROC) curve was computed which provides
a measure of discrimination performance across decision
thresholds.

\subsection{Results and Discussion}

\subsubsection{Results: Type 1 Error Analytic Formula Performance}
\begin{figure}
\centering
\includegraphics[width=0.80\linewidth]{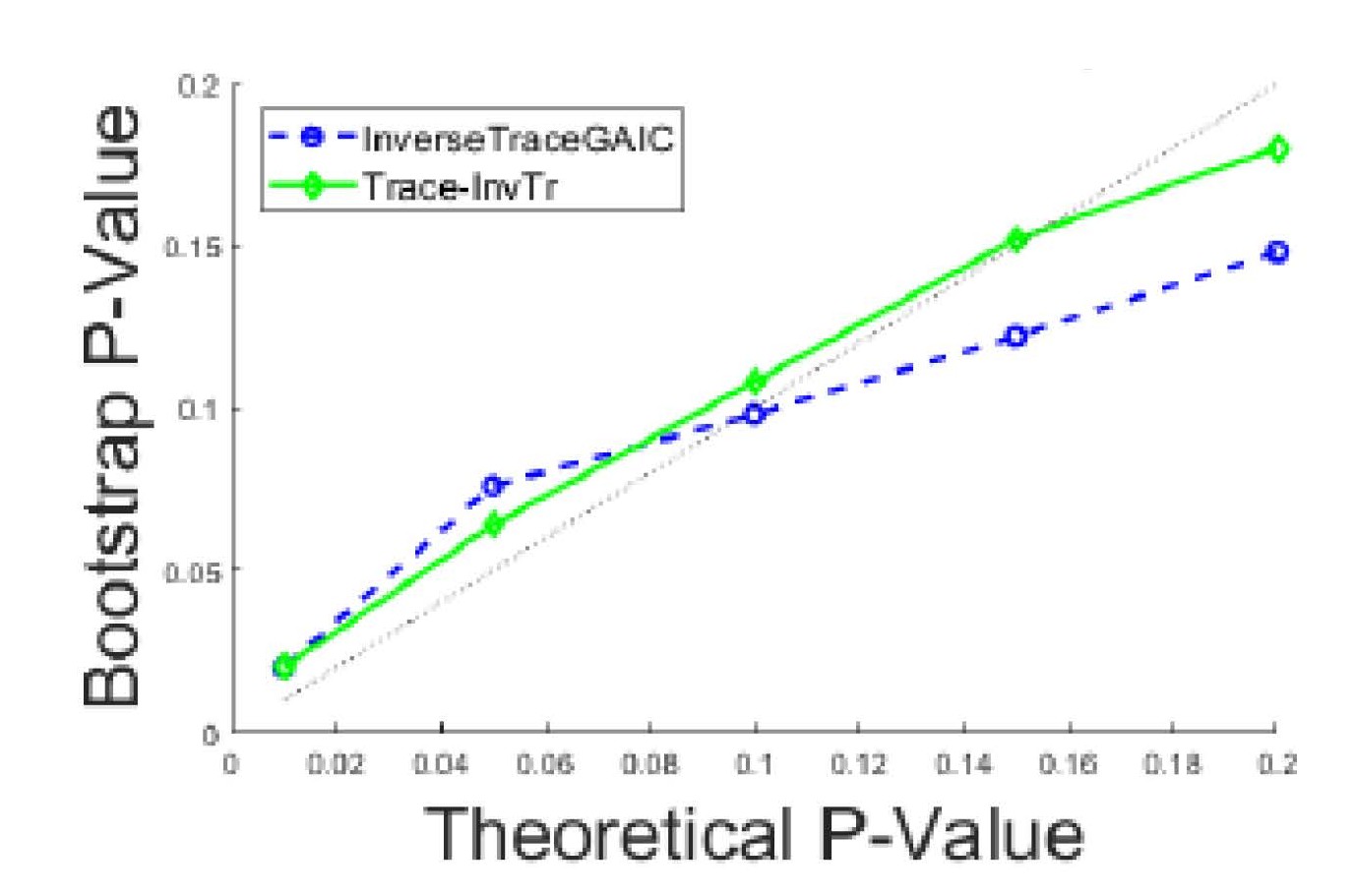}
\caption{{\bf Simulated Bootstrap P-Values versus Analytic Formula Theoretical P-Values.} The P-value
plot shows that  the Type 1 error rate is estimated by a bootstrap
simulation resampling method (Bootstrap P-Value) has good agreement
with the p-value theoretical formulas (which do not require bootstrap
simulations) for typically used p-values (e.g., p = 0.05).}
\label{fig_pvalueplot}
\end{figure}

Figure \ref{fig_pvalueplot} is a P-value plot which is designed
to evaluate the performance of the two new GIMTs for controlling for
Type 1 errors for various choices of significance levels. Figure
\ref{fig_pvalueplot} plots the Type 1 error of the Wald test in
(\ref{WaldTest}) as a function of the significance level. Given 
a sufficiently
large sample size and a sufficiently large number of simulated bootstrap
data sets, the Bootstrap and Theoretical P-values should agree.

Both the Inverse Trace GAIC
GIMT and Trace Inverse-Trace GAIC GIMT provide reasonable 
estimates of the Type 1 error performancew within the range
of significance levels typically used in practice ($ p < 0.1)$. For the
range of $p = 0.1$ through $p = 0.2$ the two-degree of freedom 
Trace Inverse-Trace GAIC GIMT formulas for computing
p-values using (\ref{WaldTest}) which have good agreement with the mathematical theory.
Although the one-degree of freedom Inverse Trace GAIC GIMT deviates
from its theoretical p-value predictions in the range of $p = 0.1$ through 
$p = 0.2$, the estimated p-values remain less than the desired Type 1 
error cutoff (i.e., significance level) so the Type 1 error rate still remains correctly controlled at the desired significance level.

\subsubsection{Results: Power Analysis using Alternative Models with Missing Skills}
\begin{figure}
\centering
\includegraphics[width=0.80\linewidth]{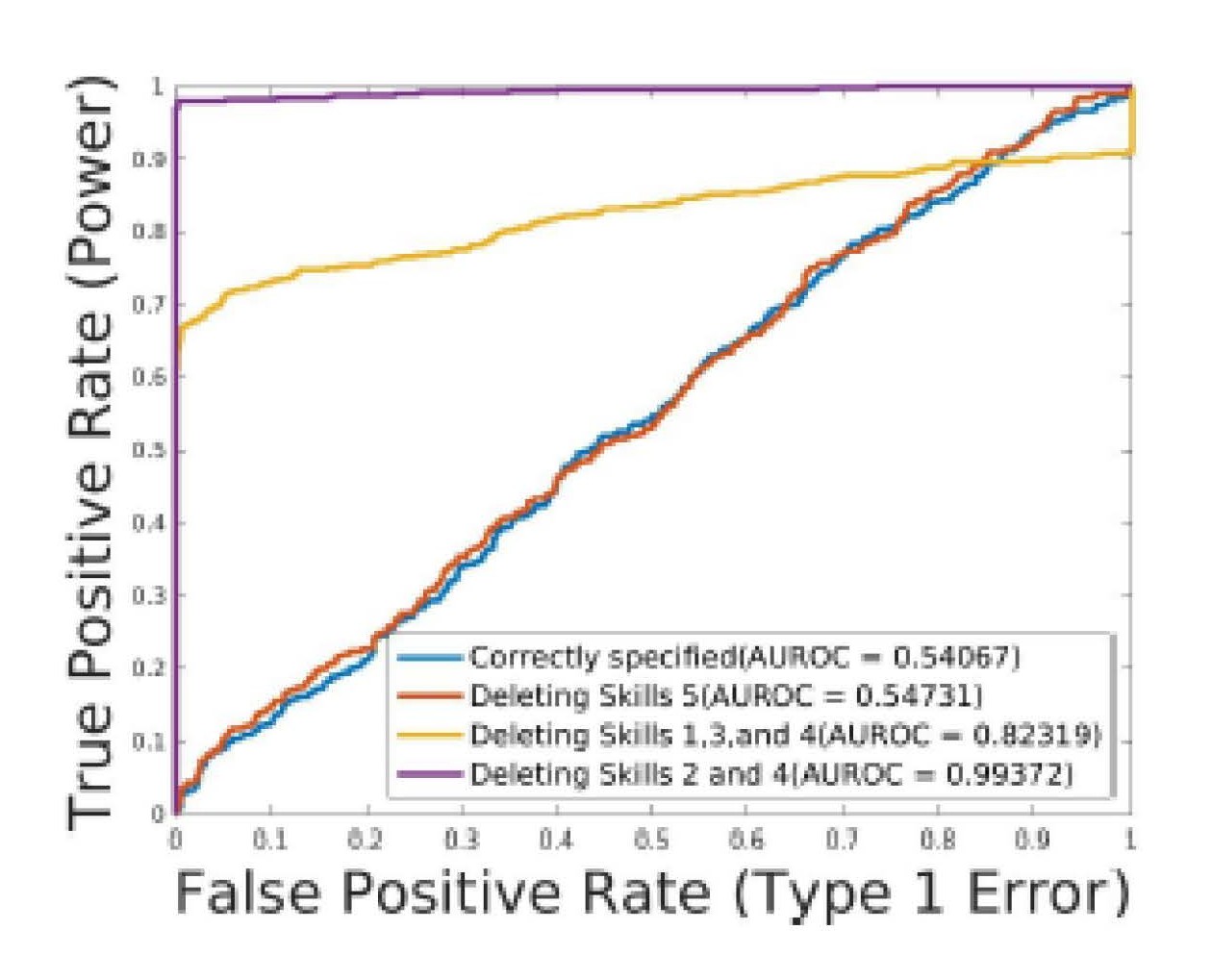}
\caption{{\bf ROC Curves for investigating Inverse-Trace GAIC GIMT discrimination performance using alternative misspecified models with deleted skills. }
Discrimination performance was excellent when the alternative model
involved deleting key skills 2 and 4. Moderate discrimination
performance was obtained by deleting skills 1, 3, and 4. The Inverse Trace
GAIC GIMT could not discriminate between the correctly specified model
and a misspecified model generated by deleting the less relevant skill 5.
}
\label{fig_invtraceroc}
\end{figure}

\begin{figure}
\centering
\includegraphics[width=0.80\linewidth]{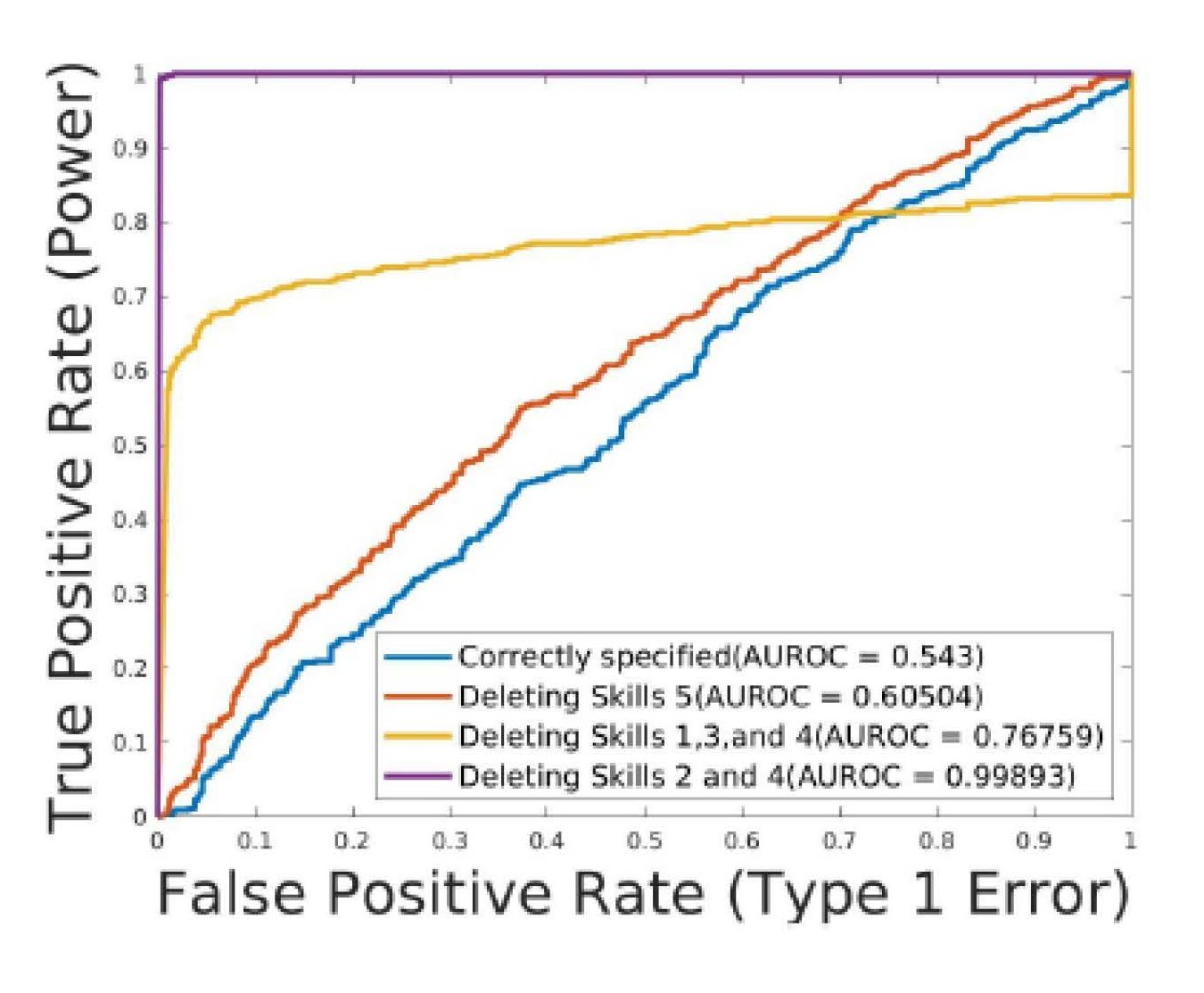}
\caption{{\bf ROC Curves for investing the Trace Inverse-Trace GAIC GIMT
discrimination performance using alternative misspecified models with
deleted skills.}
Discrimination performance was excellent when the alternative model
involved deleting key skills 2 and 4. Moderate discrimination
performance was obtained by deleting skills 1, 3, and 4. The Inverse Trace
GAIC GIMT could not discriminate between the correctly specified model
and a misspecified model generated by deleting the less relevant skill 5.
}
\label{fig_traceinvtraceroc}
\end{figure}

ROC curves generated using the methodology in Section \ref{rocmethod} are presented
in Figure \ref{fig_invtraceroc} and Figure \ref{fig_traceinvtraceroc} respectively.
In the presence of severe model misspecification associated with the removal
of key skill 2 and key skill 4, both the Inverse-Trace GAIC and Trace Inverse-Trace
GAIC show good misspecification detection performance (AUROC > 0.99). Deleting the
skills 1, 3, and 4 corresponded to a situation where the Inverse-Trace GAIC 
GIMT (AUROC = 0.82) and Trace Inverse-Trace GAIC GIMT (AUROC = 0.77) were less effective at misspecification
detection. 

When the power analysis was performed by using the correctly
specified model (no deleted skills) as the alternative hypothesis, then
misspecification performance was at chance levels (AUROC = 0.5)
for both GIMTs in Figures \ref{fig_invtraceroc} and \ref{fig_traceinvtraceroc}.
In addition, chance discrimination performance (AUROC = 0.6)
was obtained for both GIMTs when the alternative misspecified model was
generated by deleting skill 5. Inspection of the Q matrix of the original
CDM which generated the data showed that skill 5 was connected to fewer items.

\section{General Discussion}
Two entirely new tests for model misspecification for CDMs which have
never been previously proposed or published in the existing applied
statistics or psychometric literature were derived. These two
new misspecification tests are Wald tests whose test statistics
have a chi-square distribution with only 1 or 2 degrees of freedom
regardless of data set size or model complexity. The Trace Inverse-Trace
GAIC GIMT, in particular, is associated with a 2 degree of freedom
test statistic and is a non-directional GIMT.

Next, the performance of these two new GIMTs were investigated in a series
of simulation studies. First, the performance of the new analytic formulas
for estimating p-values and reported as p-value plots. Second, the
performance of the new analytic formulas for classifying models
as misspecified or correctly specified was investigated and reported
using ROC curves.

The new statistical tests showed good level (Type 1 error estimation) performance and power performance for a sample size of $n=134$. Future
work will futher investigate how the asymptotic behavior of the performance of the
CDM GIMTs changes as sample size increases.

\printendnotes

\paragraph{Author Contributions:}
Both authors contributed equally to the theoretical work, empirical work,
simulation work, and manuscript preparation.

\paragraph{Competing Interests:}
None

\printbibliography

\end{document}